\documentclass[sigconf]{acmart}

\AtBeginDocument{%
  }

\setcopyright{rightsretained}
\copyrightyear{2026}
\acmYear{2026}
\acmDOI{10.1145/XXXXXXX.XXXXXXX}
\acmConference[RECODE '26]{1st International Workshop on Code Translation, Transformation, and Modernization}{April 13, 2026}{Rio de Janeiro, Brazil}
\acmBooktitle{Proceedings of the IEEE/ACM 48th International Conference on Software Engineering: Workshops (ICSE '26), April 14--20, 2026, Rio de Janeiro, Brazil}
\acmPrice{15.00}
\acmISBN{978-1-4503-XXXX-X/26/04}

\usepackage{subcaption}     
\usepackage{xcolor}         
\definecolor{BrickRed}{rgb}{0.8, 0.25, 0.33}
\usepackage{tabularx}
\usepackage{enumitem}
\usepackage{multirow}
\usepackage{listings}
\usepackage{xcolor} 
\usepackage{lmodern} 

\begin{document}

\title{When Many-Shot Prompting Fails: An Empirical Study of LLM Code Translation}

\author{Amirkia Rafiei Oskooei}
\email{amirkia.oskooei@intellica.net}
\email{amirkia.oskooei@std.yildiz.edu.tr}
\affiliation{%
  \institution{Intellica Business Intelligence, R\&D Center}
  \city{Istanbul}
  \country{Turkey}
}
\affiliation{%
  \institution{Yildiz Technical University, Department of Computer Engineering}
  \city{Istanbul}
  \country{Turkey}
}

\author{Kaan Baturalp Cosdan}
\author{Husamettin Isiktas}
\email{{baturalp.cosdan,husamettin.isiktas}@std.yildiz.edu.tr}
\author{Mehmet S. Aktas}
\email{aktas@yildiz.edu.tr}
\affiliation{%
  \institution{Yildiz Technical University, Department of Computer Engineering}
  \city{Istanbul}
  \country{Turkey}
}

\renewcommand{\shortauthors}{Rafiei Oskooei et al.}

\begin{abstract}
Large Language Models (LLMs) with vast context windows offer new avenues for in-context learning (ICL), where providing many examples ("many-shot" prompting) is often assumed to enhance performance. We investigate this assumption for the complex task of code translation. Through a large-scale empirical study of over 90,000 translations, we systematically evaluate the impact of scaling in-context examples from zero-shot to many-shot configurations of up to 625 examples, with prompts spanning from ~100K to ~800K tokens. Our findings reveal a "many-shot paradox": while static similarity metrics may modestly improve with more examples, functional correctness consistently peaks with few-shot prompting (5-25 examples). Providing substantially more examples often degrades this crucial functional performance. This study highlights that for code translation, the quality of a few well-chosen examples outweighs sheer quantity, challenging the universal efficacy of "more is better" for ICL and underscoring the task-dependent nature of optimal prompting strategies. Our results have significant implications for effectively leveraging LLMs in software engineering.
\end{abstract}



\begin{teaserfigure}
  \centering
  \includegraphics[width=\textwidth]{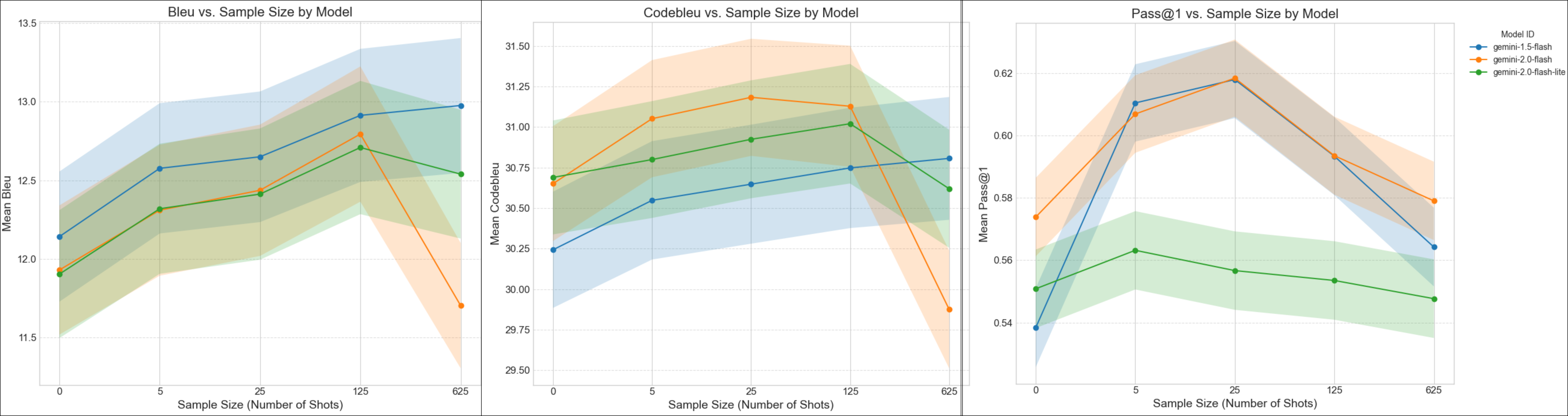}
  \caption{
    The "many-shot paradox" in code translation, illustrated across static and dynamic metrics. 
    From left to right: \textbf{(a) BLEU} and \textbf{(b) CodeBLEU} show that static similarity improves from zero-shot to few-shot, but performance plateaus or even declines in the many-shot regime. The paradox is most evident in \textbf{(c) Pass@1}, where functional correctness, measured by compilation success, consistently peaks in the few-shot range (5-25 examples) before significantly degrading for all models as the number of examples increases. This demonstrates that for code translation, more context does not equate to better functional outcomes.
  }
  \label{fig:teaser}
\end{teaserfigure}


\maketitle

\section{Introduction}
\label{sec:introduction}

The advent of Large Language Models (LLMs) with million-token context windows has unlocked new frontiers for in-context learning (ICL). This paradigm allows models to adapt to new tasks at inference time by conditioning on demonstration examples provided directly in the prompt. A growing body of research suggests that scaling the number of these examples into the hundreds or thousands---a technique known as "many-shot" prompting---leads to performance gains across diverse domains, often matching or exceeding the capabilities of fine-tuned models \cite{agarwal2024many,lampinen2025generalization,jiang2024many}. This has fueled a compelling narrative: for LLMs, more context is better.

However, code translation, a critical software engineering task for modernizing legacy systems and ensuring cross-platform compatibility, presents a unique challenge. Unlike many natural language tasks, successful code translation is not merely a matter of lexical or syntactic similarity. It demands a deep semantic understanding to produce code that is functionally equivalent---it must compile, execute correctly, and produce the exact same output as the source. This strict requirement for functional correctness makes it a demanding testbed for ICL.

This paper investigates a critical and underexplored question: does the "more examples are better" mantra hold for the intricate and logically constrained task of code translation? We present one of the first large-scale empirical studies to address this, analyzing over 90,000 translation experiments across 30 language pairs using Google's Gemini model family. Our findings reveal a "many-shot paradox": while static similarity metrics may modestly improve with more examples, functional correctness consistently peaks with a surprisingly small number of examples (5--25 shots). Providing substantially more examples often degrades this crucial functional performance, challenging the universal efficacy of many-shot prompting for all tasks. The key contributions of this work are:
\begin{itemize}[nosep]
    \item Empirical evidence of a "many-shot paradox" in code translation, where functional correctness degrades with excessive examples.
    \item Identification of an optimal "few-shot" range (5--25 examples) that maximizes functional performance and cost-effectiveness.
    \item An analysis showing a divergence between static similarity metrics and dynamic, execution-based correctness.
\end{itemize}
\section{Related Works}
\label{sec:related_works}

Our research is situated at the intersection of LLMs for code translation and the dynamics of in-context learning (ICL), particularly in the many-shot regime.

\paragraph{LLMs for Code Translation}
The application of LLMs to software engineering has grown substantially, with code translation being a prominent use case. Researchers have developed advanced, neuro-symbolic frameworks like AlphaTrans for repository-level translation, demonstrating the potential for complex, large-scale migrations \cite{ibrahimzada2024alphatrans}. However, the reliability of LLM-generated code remains a challenge, as models frequently introduce subtle bugs, including compilation errors, runtime exceptions, and functional deviations \cite{pan2024lost}. This has spurred research into improving the inputs to the LLM. Recognizing that models often overlook critical syntactic information, recent work has explored using ICL to post-incorporate code structural knowledge via Abstract Syntax Tree (AST) coverage, improving translation quality without retraining the model \cite{du_post-incorporating_2025}. These studies highlight that while LLMs are powerful, achieving high-fidelity code translation requires careful prompting and a focus on semantic correctness.

\paragraph{Many-Shot In-Context Learning}
With the advent of models supporting long context windows, research has expanded from few-shot to "many-shot" ICL. Foundational studies have reported significant benefits from this approach, finding that scaling from few-shot to many-shot ICL leads to substantial performance gains across a wide variety of tasks, even matching fine-tuning performance in some cases \cite{agarwal2024many,lampinen2025generalization}. This trend has been observed in other domains as well; for example, in multimodal foundation models, performance continues to improve log-linearly with thousands of examples \cite{jiang2024many}. This line of work has pushed the boundaries of ICL to an extreme scale, showing that for datasets with large label spaces, performance can continue to increase, reinforcing the "more is better" paradigm \cite{bertsch_-context_2025}.

\paragraph{Positioning Our Work}
While many-shot ICL often benefits NLP tasks, emerging work suggests performance can plateau or decline due to noise \cite{zhang2025more}. We bridge this gap by systematically investigating this phenomenon in code translation. We provide the first large-scale evidence of a "many-shot paradox" specifically for code, demonstrating via dynamic metrics that naive scaling of examples can degrade functional correctness in this structurally sensitive domain.
\section{Methodology}
\label{sec:methodology}
\begin{figure}[t]
    \centering
    \includegraphics[width=0.8\columnwidth]{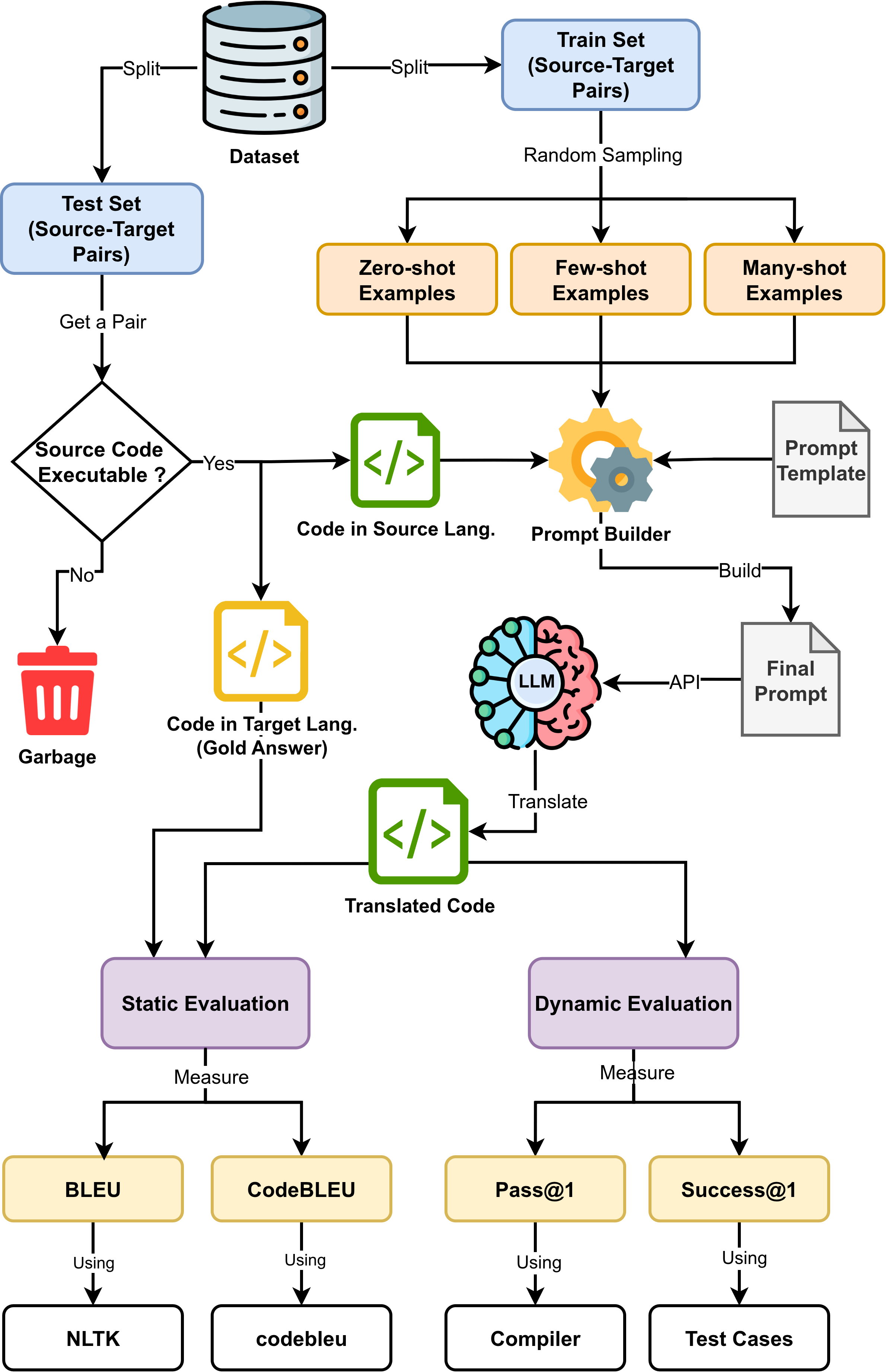} 
    \caption{An overview of our experimental methodology.}
    \label{fig:methodology_diagram}
\end{figure}

Our methodology is designed to provide a rigorous, large-scale empirical investigation into the efficacy of in-context learning (ICL) for code translation, specifically addressing the following questions:
\begin{itemize}[nosep]
    \item[\textbf{RQ1:}] How does scaling from zero-shot to many-shot in-context examples affect both the syntactic quality and, more importantly, the functional correctness of LLM-based code translation?
    \item[\textbf{RQ2:}] What is the optimal range of in-context examples, and how do scaling effects vary across LLMs within the same model family?
\end{itemize}
Figure~\ref{fig:methodology_diagram} provides a high-level overview of our experiments.

\subsection{Models}
We evaluate three variants from Gemini family: \textit{Gemini 1.5 Flash}, \textit{Gemini 2.0 Flash}, and \textit{Gemini 2.0 Flash Lite} \cite{team2023gemini,team2024gemini}. This selection was strategically motivated by several critical factors for our study's validity:\footnote{At the time of our experiments, publicly available LLMs with million-token context windows were scarce. The Gemini family was a primary option and had been adopted in foundational many-shot ICL studies \cite{agarwal2024many,lampinen2025generalization}, making our results directly comparable. This choice allows us to isolate the effect of prompting strategy on a consistent architecture.}
\begin{enumerate}[nosep]
    \item \textbf{Long Context}: All selected models support context windows of at least 1 million tokens, which is a prerequisite for investigating many-shot prompting at the scale of 625 examples.
    \item \textbf{Controlled Comparison}: By testing variants within the same model family, we can isolate the impact of model scale on ICL efficacy while holding the core architecture constant, avoiding confounding factors from cross-architecture comparisons.
    \item \textbf{Research Continuity}: Our choice of Gemini aligns with the foundational many-shot ICL studies we are in dialogue with, enabling a direct comparison of our code translation findings against results from other NLP and reasoning tasks \cite{agarwal2024many,lampinen2025generalization}.
\end{enumerate}

\subsection{Dataset and Prompting Configurations}
Our empirical study is grounded in the \textbf{CodeTransOcean} benchmark \cite{yan2023codetransocean}. We focus on six diverse programming languages (C, C++, C\#, Java, Go, Python), resulting in 30 distinct translation pairs. For each pair, we randomly sampled 200 unique source code snippets from a held-out test partition, leading to 90,000 total translation experiments. Each task includes a robust test suite of input-output pairs covering edge cases to ensure strict functional verification. Key dataset properties are summarized in Table~\ref{tab:dataset_summary}.

We investigate the full spectrum of ICL through five shot configurations: \textbf{zero-shot} (0 examples), \textbf{few-shot} (5 and 25 examples), and \textbf{many-shot} (125 and 625 examples). For all configurations with demonstrations, the examples were randomly sampled from a disjoint training partition to prevent data leakage. This random selection serves as a baseline to isolate the effect of quantity, though it may introduce noise compared to curated examples. Our prompt was deliberately minimal to test the models' core ICL capabilities, containing only a task instruction, output format guidance, the N in-context examples, and the source code to be translated.

\begin{table}[t]
\centering
\caption{Summary of the Dataset and Experimental Setup.}
\label{tab:dataset_summary}
\scriptsize
\begin{tabularx}{\columnwidth}{@{}ll@{}}
\toprule
\textbf{Property} & \textbf{Description} \\
\midrule
\textbf{Benchmark} & CodeTransOcean \cite{yan2023codetransocean} \\
\textbf{Languages} & C, C++, C\#, Java, Go, Python \\
\textbf{Language Pairs} & 30 (All-to-all, non-identity) \\
\textbf{Tests per Pair} & 200 unique code snippets \\
\textbf{Total Tests} & 6,000 per model \& shot config \\
\textbf{Total Translations} & 90,000 (6k $\times$ 5 configs $\times$ 3 models) \\
\bottomrule
\end{tabularx}
\end{table}

\subsection{Evaluation Framework}
Evaluating code translation requires a multifaceted approach, as syntactic similarity does not guarantee functional equivalence. Our framework integrates both static and dynamic metrics. All dynamic tests were conducted within isolated Docker containers to ensure a consistent and reproducible execution environment.

\paragraph{Static Evaluation} We use two standard metrics for an initial assessment of lexical and syntactic similarity. \textbf{BLEU} measures n-gram overlap, while \textbf{CodeBLEU} is a more sophisticated metric that incorporates syntactic information.

\paragraph{Dynamic Evaluation} The core of our evaluation is a hierarchical pipeline that assesses true functional correctness. Each translation $T$ of a source snippet $S$ with test cases $\mathcal{T}_S$ is categorized based on the first point of failure, as defined in Equation~\ref{eq:outcome}.

\begin{equation}
\label{eq:outcome}
\small
\text{Outcome}(T, S) =
\begin{cases}
\text{Compilation Error} & \text{if } \neg \mathcal{E}_{Comp}(T) \\
\text{Runtime Error} & \text{if } \mathcal{E}_{Comp}(T) \land \neg \mathcal{E}_{Run}(T|\mathcal{T}_S) \\
\text{Functional Error} & \text{if } \mathcal{E}_{Comp}(T) \land \mathcal{E}_{Run}(T|\mathcal{T}_S) \\
& \quad \land \neg \mathcal{E}_{Func}(T, S|\mathcal{T}_S) \\
\text{Successful Translation} & \text{otherwise}
\end{cases}
\end{equation}
where the outcome is determined by these events:
\begin{itemize}[nosep]
    \item $\mathcal{E}_{Comp}(T)$: The translation $T$ compiles successfully.
    \item $\mathcal{E}_{Run}(T|\mathcal{T}_S)$: $T$ executes on all test cases without runtime errors.
    \item $\mathcal{E}_{Func}(T, S|\mathcal{T}_S)$: The outputs of $T$ are semantically equivalent to the outputs of $S$ on all test cases.
\end{itemize}
From these outcomes, we derive our key dynamic metrics. \textbf{Pass@1} measures the rate of compilation success, while the stricter \textbf{Success@1} requires a translation to be fully successful (i.e., outcome is "Successful Translation").
\section{Results and Analysis}
\label{sec:results}

Our empirical findings reveal a clear and consistent "many-shot paradox" in LLM-based code translation. This section details this core finding by first examining the divergence between static and dynamic performance metrics, then analyzing the underlying shift in error distributions, and finally assessing the robustness of our findings across different language pair characteristics.

\subsection{The Many-Shot Paradox: Functional Correctness Peaks with Few Shots}
\label{sec:paradox_finding}

A central finding of our study is the stark divergence between how static similarity and dynamic functional correctness scale with an increasing number of in-context examples. This phenomenon is illustrated in our teaser figure (Figure~\ref{fig:teaser}) and detailed with full statistical breakdowns in Table~\ref{tab:merged_metrics}.

The paradox is most evident in the Pass@1 metric (Figure~\ref{fig:teaser}c), which measures compilation success. Across all three Gemini models, Pass@1 performance exhibits a distinct non-monotonic trend. It increases significantly from a zero-shot baseline to a peak within the few-shot regime (5 to 25 examples). For instance, both Gemini 1.5 Flash and 2.0 Flash reach their peak Pass@1 score of 61.8\% at 25 shots, as shown in Table~\ref{tab:merged_metrics}. Beyond this optimal range, performance consistently and significantly degrades, with Pass@1 scores at 625 shots falling back to near zero-shot levels for some models.

In contrast, the trends for static metrics are less uniform and do not reflect this clear degradation in functional viability. While BLEU (Figure~\ref{fig:teaser}a) and CodeBLEU (Figure~\ref{fig:teaser}b) scores generally improve from zero-shot to few-shot, their behavior in the many-shot regime is inconsistent. For Gemini 1.5 Flash, static scores continue to climb modestly up to 625 shots. However, for the more powerful Gemini 2.0 Flash, both BLEU and CodeBLEU scores decline after peaking at 125 and 25 shots, respectively (see Table~\ref{tab:merged_metrics}).

This divergence is critical: it indicates that in many-shot configurations, models may be learning to replicate superficial lexical or syntactic patterns from the examples, which can sometimes improve static scores, but this comes at the expense of generating structurally sound and compilable code. The consistent decline in Pass@1 suggests that for the complex task of code translation, naively increasing the quantity of examples introduces noise or conflicting signals that impair the model's ability to maintain functional correctness.

\begin{table*}[t]
\centering
\caption{Performance metrics (BLEU, CodeBLEU, Pass@1) across models and shot-sizes.}
\label{tab:merged_metrics}
\scriptsize
\renewcommand{\arraystretch}{0.95}
\setlength{\tabcolsep}{1.5pt}
\begin{tabular}{l|ccc|ccc|ccc}
\toprule
\textbf{Shots} & \multicolumn{3}{c|}{\textbf{BLEU}} & \multicolumn{3}{c|}{\textbf{CodeBLEU}} & \multicolumn{3}{c}{\textbf{Pass@1 (\%)}} \\
& \textbf{1.5 Flash} & \textbf{2.0 Flash} & \textbf{2.0 Lite} & \textbf{1.5 Flash} & \textbf{2.0 Flash} & \textbf{2.0 Lite} & \textbf{1.5 Flash} & \textbf{2.0 Flash} & \textbf{2.0 Lite} \\
\midrule
0 & 12.14 & 11.93 & 11.90 & 30.24 & 30.65 & 30.69 & 53.8 & 57.4 & 55.1 \\
5 & 12.58 & 12.31 & 12.32 & 30.55 & 31.05 & 30.80 & 60.4 & 61.5 & 56.3 \\
25 & 12.65 & 12.44 & 12.41 & 30.65 & 31.18 & 30.92 & 61.8 & 61.8 & 55.9 \\
125 & 12.91 & 12.79 & 12.71 & 30.75 & 31.13 & 31.02 & 60.2 & 60.1 & 55.6 \\
625 & 12.98 & 11.70 & 12.54 & 30.81 & 29.87 & 30.62 & 56.4 & 57.9 & 54.8 \\
\bottomrule
\end{tabular}
\end{table*}

\subsection{Error Distribution Analysis}
\label{sec:error_analysis}

\begin{figure*}[t]
    \centering
    \includegraphics[width=1.0\textwidth]{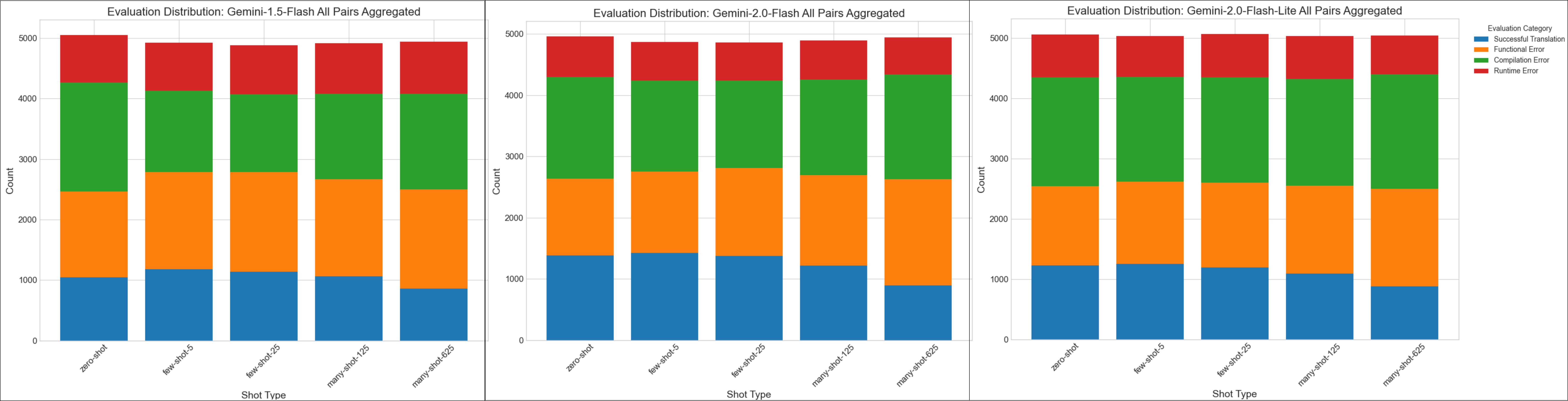}
    \caption{Distribution of evaluation outcomes (aggregated across all language pairs) for each Gemini model variant. Moving from zero-shot to few-shot (5-25) reduces Compilation Errors and increases Successful Translations. In many-shot settings, however, successful translations decline while Functional and Runtime Errors become more prevalent.}
    \label{fig:error_distributions}
\end{figure*}

To understand \textit{why} functional correctness degrades in many-shot settings, we analyzed the distribution of translation outcomes. Figure~\ref{fig:error_distributions} illustrates the shift in error types for all three Gemini models as the number of in-context examples increases. Moving from zero-shot to few-shot (5-25 examples) provides a clear benefit: a significant reduction in Compilation Errors (CE). This indicates that a small number of examples is highly effective at guiding the models to produce syntactically valid code. For Gemini 2.0 Flash, this initial guidance also leads to the highest proportion of Successful Translations (ST), our strictest metric for functional equivalence.

However, this positive trend does not sustain. As we scale to many-shot configurations (125 and 625 shots), the rate of Successful Translations stagnates and, for all models, eventually declines from its peak. Critically, this decline is accompanied by a rise in Functional Errors (FE) and Runtime Errors (RE), which become the dominant failure modes. This suggests that while ICL can effectively teach syntactic patterns, achieving deep semantic equivalence and avoiding subtle runtime bugs is a more profound challenge that is not solved---and is sometimes exacerbated---by a brute-force increase in examples. The overabundance of context appears to shift the challenge from syntax to semantics, where the models struggle more.

\subsection{Influence of Language Characteristics}
\label{sec:language_influence}

\begin{figure}[h!]
    \centering
    \includegraphics[width=0.8\columnwidth]{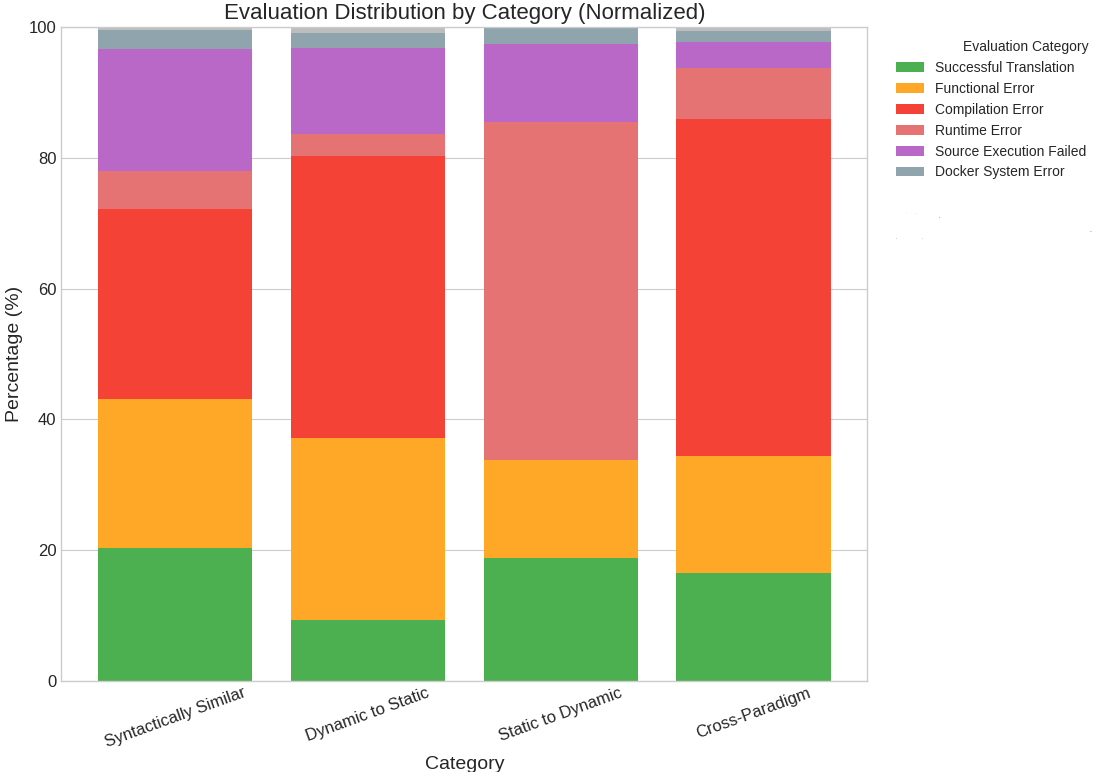}
    \caption{Outcome distribution by category; syntactically similar pairs yield higher success rates.}
    \label{fig:category_distribution}
\end{figure}

To assess the generalizability of our findings, we analyzed performance across four categories of language pairs based on their structural and semantic properties. Table~\ref{tab:category_metrics_full} presents a detailed breakdown for representative pairs, showing a clear performance hierarchy. \textit{Syntactically Similar} languages (e.g., Java to C\#) consistently achieve the highest success rates, while translations involving disparate paradigms or type systems prove more challenging. The distribution of outcomes, shown in Figure~\ref{fig:category_distribution}, reinforces this finding, with syntactically similar pairs yielding the highest proportion of successful translations.

\begin{figure*}[h!]
    \centering
    \includegraphics[width=1.0\textwidth]{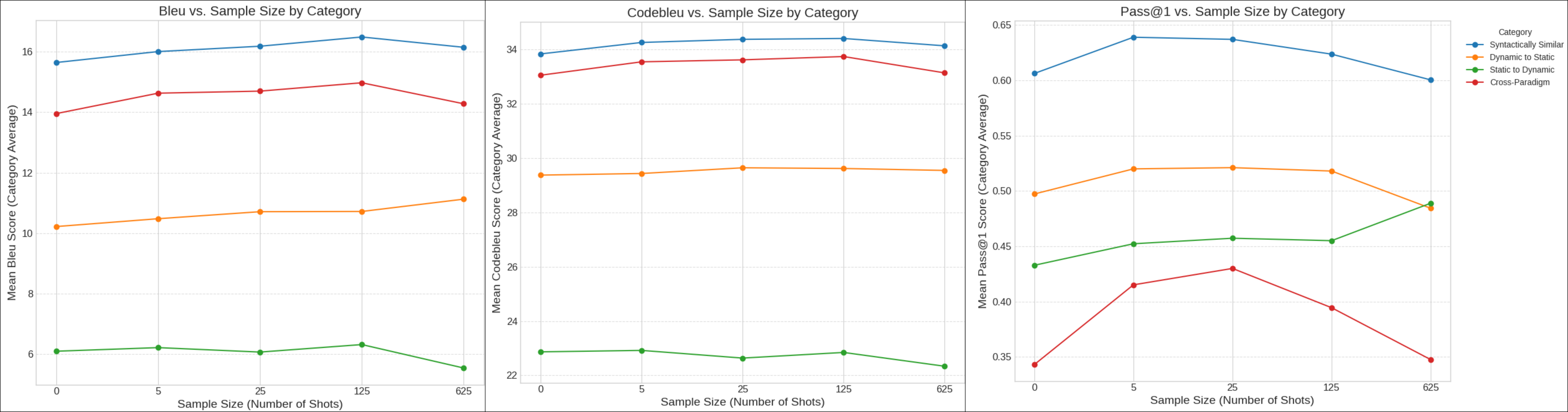}
    \caption{Performance metrics broken down by language pair category. The non-monotonic trend in Pass@1 (rightmost plot) is observed across three categories, demonstrating the robustness of the "many-shot paradox."}
    \label{fig:category_plots}
\end{figure*}

Crucially, despite these baseline performance differences, the many-shot paradox persists across all categories. Figure~\ref{fig:category_plots} shows that the non-monotonic Pass@1 curve is a consistent phenomenon, regardless of the translation's inherent difficulty. Even for the easiest category of syntactically similar pairs, Pass@1 performance peaks in the few-shot range and declines thereafter. This reinforces our central conclusion: while linguistic distance is a major factor in overall translation difficulty, the degradation of functional correctness in many-shot settings appears to be a fundamental behavior of ICL for the task of code translation, not an artifact of specific challenging language pairs.

\begin{table}[t]
\centering
\caption{Static and dynamic metrics for representative language pairs within each category, aggregated across all models and shot sizes. The success rate shows the raw count of successful translations over the total number of valid attempts.}
\label{tab:category_metrics_full}
\scriptsize
\begin{tabular}{|l|c|c|c|c|}
\hline
\textbf{Language Pair} & \textbf{BLEU} & \textbf{CodeBLEU} & \textbf{pass@1} & \textbf{Success/Total} \\
\hline
\multicolumn{5}{|l|}{\textbf{Syntactically Similar Pairs}} \\
\hline
C\# $\rightarrow$ C++ & 13.11 & 32.07 & 0.97 & 641 / 2887 \\
C++ $\rightarrow$ Java & 19.41 & 35.49 & 0.55 & 634 / 2863 \\
Java $\rightarrow$ C\# & 15.78 & 35.01 & 0.34 & 543 / 2934 \\
\hline
\multicolumn{5}{|l|}{\textbf{Dynamically Typed to Statically Typed Pairs}} \\
\hline
Python $\rightarrow$ C++ & 7.80 & 27.06 & 0.97 & 475 / 2899 \\
Python $\rightarrow$ Go & 12.77 & 31.74 & 0.08 & 93 / 2950 \\
Python $\rightarrow$ Java & 11.42 & 29.78 & 0.48 & 268 / 2866 \\
\hline
\multicolumn{5}{|l|}{\textbf{Statically Typed to Dynamically Typed Pairs}} \\
\hline
C\# $\rightarrow$ Python & 8.13 & 23.11 & 0.49 & 511 / 2916 \\
C++ $\rightarrow$ Python & 5.05 & 22.52 & 0.43 & 632 / 2924 \\
Java $\rightarrow$ Python & 4.98 & 22.53 & 0.45 & 547 / 2930 \\
\hline
\multicolumn{5}{|l|}{\textbf{Cross-Paradigm Pairs}} \\
\hline
C $\rightarrow$ Java & 15.52 & 32.40 & 0.59 & 641 / 2909 \\
Go $\rightarrow$ C\# & 12.09 & 33.95 & 0.45 & 686 / 2933 \\
Java $\rightarrow$ Go & 15.93 & 33.86 & 0.12 & 159 / 2956 \\
\hline
\end{tabular}
\end{table}

\subsection{Cost-Performance Analysis}
Beyond the degradation in functional correctness, the many-shot paradox also has significant economic implications. As shown in Table~\ref{tab:prompt_statistics}, the prompt lengths and the corresponding API costs grow exponentially as the number of in-context examples increases. This analysis highlights the stark trade-off between the theoretical appeal of providing more context and the practical realities of cost and performance.
For our best-performing model, Gemini 2.0 Flash, moving from the peak-performance 25-shot configuration to the 625-shot configuration represents a more than 21-fold increase in the estimated cost per translation. Crucially, this substantial financial and computational investment does not correspond with an improvement, but with a tangible degradation in functional correctness, as demonstrated in our Pass@1 and error distribution analyses. This finding provides evidence that for the task of code translation, a well-calibrated few-shot approach is not only the most effective strategy but also the more economically viable, rendering large-scale many-shot prompting both technically suboptimal and financially inefficient.

\begin{table}[h]
\centering
\caption{Estimated prompt lengths (tokens) and API costs (USD) per translation, assuming an average output of 1000 tokens. (Google AI Studio pricing)}
\label{tab:prompt_statistics}
\scriptsize
\begin{tabular}{|c|c|c|c|}
\hline
\textbf{Model} & \textbf{Config} & \textbf{Avg. Length} & \textbf{Avg. Price} \\
\hline
\multirow{5}{*}{gemini-1.5-flash} 
& zero-shot & 63 & \$0.0091 \\
& few-shot-5 & 4441 & \$0.0190 \\
& few-shot-25 & 22011 & \$0.0585 \\
& many-shot-125 & 108156 & \$0.2524 \\
& many-shot-625 & 553738 & \$1.2549 \\
\hline
\multirow{5}{*}{gemini-2.0-flash} 
& zero-shot & 65 & \$0.0122 \\
& few-shot-5 & 4452 & \$0.0254 \\
& few-shot-25 & 22101 & \$0.0783 \\
& many-shot-125 & 108658 & \$0.3380 \\
& many-shot-625 & 556218 & \$1.6807 \\
\hline
\multirow{5}{*}{gemini-2.0-flash-lite} 
& zero-shot & 65 & \$0.0091 \\
& few-shot-5 & 4452 & \$0.0190 \\
& few-shot-25 & 22101 & \$0.0587 \\
& many-shot-125 & 108658 & \$0.2535 \\
& many-shot-625 & 556218 & \$1.2605 \\
\hline
\end{tabular}
\end{table}
\section{Conclusion}
\label{sec:conclusion}

This study challenges the universality of the "more is better" assumption for in-context learning. While extensive many-shot prompting has demonstrated clear benefits in retrieval and reasoning tasks, our analysis of code translation identifies a divergent "many-shot paradox," where functional correctness consistently peaks in the few-shot regime (5–25 examples) and degrades with larger contexts. This aligns with a growing body of literature suggesting that for cognitively demanding tasks, LLMs operate within a finite "attention budget," where excessive or less relevant context can dilute focus and introduce noise rather than reinforcing patterns \cite{liu2023lost,shi2023large,oskooei2025natural}. Practically, our results suggest prioritizing the curation of small, high-quality exemplar sets, which is not only more effective but also reduces API costs by 20-30x compared to many-shot prompting. Future work must broaden this analysis to other architectures (e.g., Llama) and move beyond our random sampling baseline to test relevance. We hypothesize that "high-quality" examples—idiomatic and semantically related—may mitigate the paradox, whereas our single-pass, random approach exposes the noise sensitivity of current models. Developing iterative, feedback-driven workflows may further overcome these limitations.

\begin{acks}
We gratefully acknowledge Intellica Business Intelligence Consultancy for providing the resources and research environment at their R\&D Center to conduct this study. 
\end{acks}

\bibliographystyle{ACM-Reference-Format}
\bibliography{references}



\end{document}